\documentclass[12pt]{article}
\usepackage{graphicx}
\usepackage[cp1251]{inputenc}
\begin{document}
\pagestyle{empty}

\pagestyle{headings} {\title{Solar neutrinos as indicators of the
Sun's activity}}
\author{O.M.Boyarkin$^a$\thanks{E-mail:oboyarkin@tut.by},
\ I.O.Boyarkina\thanks{E-mail:estel20@mail.ru}\ $^b$\\
$^a$\small{\it{Belarusian State University,}}\\
\small{\it{Dolgobrodskaya Street 23, Minsk, 220070, Belarus}}\\
$^b$\small{\it{University of Rome Tor Vergata}}\\
\small{\it{Via Orazio Raimondo 18, Roma 00173 Lazio, Rome,
Italy}}}
\date{}
\maketitle
\begin{abstract}
Opportunity of the solar flares (SF's) prediction observing the
solar neutrino fluxes is investigated. In three neutrino
generations the evolution of the neutrino flux traveling the
coupled sunspots (CS's) which are the SF source is considered. It
is assumed that the neutrinos possess both the dipole magnetic
moment and the anapole moment while the magnetic field above the
CS's may reach the values $10^5-10^6$ Gs, displays the twisting nature
and has the nonpotential character. The possible resonance conversions of
the solar neutrino flux are examined. Since the $\nu_{eL}\to \nu_{\mu L}$
resonance takes
place before the convective zone, its existence can in no way be
connected with the SF. However, when the solar neutrino flux moves
through the CS's in the preflare period, then it may
undergo the additional resonance conversions and, as a result,
depleting the electron neutrinos flux may be observed.
\\[1mm]

PACS number(s): 12.60.Cn, 14.60.Pg, 96.60.Kx, 95.85.Qx, 96.60.Rd.
\end{abstract}

Keys words: Coupled sunspots, magnetic reconnection model, solar flares,
neutrino, dipole magnetic moment, anapole moment, neutrino telescopes,
resonance transitions, $\nu_e$-induced $\beta$-decays.

\section{Introduction}

At certain conditions the evolution of active regions on the Sun
may lead to the appearance of solar flares (SF's) that occur in
the solar atmosphere and release enormous amounts of energy, over
the entire electromagnetic spectrum. The energy generated during
the SF is about of $10^{28}-10^{33}$ erg. Moreover, as it was
shown in Ref. \cite{ML17}, the super-SF's with energy as large as
$10^{36}$ erg are also possible. This gigantic energy is released
on the Sun in a few minutes and corresponds to an average power of
$\mbox{few}\times10^{29}$ erg/s. However, this is less than
hundredths portion of a percent of the total solar radiation power
in the optical range which is equal to $4\times10^{33}$ erg/s.

The SF's are quite prominent in X-rays, UV, and optical lines and
they are often (but not always) followed by eruptions that throw
out solar coronal plasma into the interplanetary space (coronal
mass ejections - CME's). In relation to their peak X-ray
intensity, as recorded by the National Oceanic and Atmospheric
Administration's Geostationary Operational Environmental Satellite
system, flares are separated into classes, the strongest and most
important being X, M and C (in decreasing order). Flare
classification is logarithmic, with a base of 10, and is
complemented by decimal sub-classes (e.g. M5.0, C3.2 etc.).

It should be noted that flare events also occur in other
first-generation stars. Remember, first generation stars consist
of only from ingredients provided directly by the big bang,
namely, essentially from hydrogen and helium. Therefore, the study
of the SF's sheds light on the structure and evolution of the
Universe. Our comprehension of the SF's has been greatly enhanced
in recent times, both from a theoretical and observational
viewpoint \cite{AOB}. These achievements have been supplemented by
a great deal of data from the Kepler mission \cite{SC14}, which
surveyed $\sim10^5$ stars of M-, K-, and G-types and produced
detailed statistics concerning the frequency of large flares with
energies of order $10^{33}$ erg.

The high-power SF's can be especially destructive when they are
aimed towards the direction of the Earth. They cause problems with
power grids, radio blackouts on Earth, mutations in DNA,
destruction of ecosystems, breakdowns of different instruments on
the satellites and so on.

The strongest observed SF and accompanying CME was the Carrington
event that took place in 1859. It was about twice as big as the
strongest events observed during the space era. The SF which has
occurred at 4 August 1972 led to the triggering of magnetic
detonators of American underwater mines in the vietnamese port Hon
La. The SF's and their accompanying CME's which have taken place
between mid-October to early November 2003 peaking around 28 and
29 October (so called Halloween solar storms) even caused failure
to the power supply of the Japanese Earth-resource satellite, the
Advanced Earth Observation Satellite-II "Midori II", and made it
inoperative, while the effects of the Halloween solar storms
extended beyond the Earth to Mars and caused the Mars Odyssey
spacecraft to go into deep safe-mode \cite{REL}.

Therefore, for our increasingly technologically dependent society
it is of great practical significance to predict when and how
large the SF's will be. Previous studies on predicting solar
eruptive phenomena mainly employed measurements of the active
region (AR) magnetic field in the solar photosphere to calculate
the physical indices of the AR's and connect these indices to the
occurrences of the SF's and CME's \cite{RLM}. These SF prediction
is mainly fulfilled by using space-borne instruments such as the
Atmospheric Imaging Assembly, the Helioseismic and Magnetic Imager
on the Solar Dynamics Observatory, the Large Angle Spectroscopic
Coronagraph on the Solar and Heliospheric Observatory, and
Geostationary Operational Environmental Satellite series. However,
it does not mean that the ground-based telescopes are no longer
useful for the SF's prediction. There are several observational
methods from the ground such as a coronagraph, a magnetogram, a
continuum light observation, and a H-alpha observation. For
example, at Hida Observatory in Kyoto University, there is a
powerful instrument observing the Sun in a H-alpha line and its
wings called Solar Dynamics Doppler Imager installed on Solar
Magnetic Activity Research Telescope \cite{KI17}.

The $\gamma$-telescopes observing the Sun continuously collect
electromagnetic and particle measurements related to SF's, CMEs
and this huge amount of observations  must be transferred, stored,
and handled. To deal with these large amount of solar observation
data, a new method of Big Data Mining, also called Machine
Learning (ML) has been developed. The ML method has been using
different models, such as support vector machines \cite{RQ07},
neural networks \cite{OWA}, a regression model \cite{JYL}, an
extremely randomized trees \cite{NN17} and so on. An introduction
to ML research can be found in several textbooks (see, for
example, \cite{TH09}). The ML can clarify which feature is most
effective for predicting the SF's. However, it is still not clear
which model is the best for prediction in an operational setting.

However, the Sun radiates not only photons by which we could
define its state, the Sun is also a powerful source of neutrinos.
In the result of thermonuclear fusion reactions in the Sun's core
the total electron neutrino flux falling on a terrestrial surface could be
as large as $\Phi_{\nu}\simeq6\times10^{10}\
\mbox{cm}^{-2}\mbox{s}^{-1}.$ For the first time a correlation of
a neutrino flux with the SF's was predicted in Ref's.
\cite{OBD95,OM95}. Later this hypothesis has received support
through experiments which have demonstrated decreasing the
$\beta$-decay rate of some elements of the periodic table during
the SF's \cite{JHJ2009,DEK12,DOK13,TM16,PA18}. Early result was
presented by Jenkins and Fischbach \cite{JHJ2009} who have
detected this decreasing for $^{54}\mbox{Mn}$ at the level of
$\sim7\sigma$ before the large SF which was at 2006 Dec.13. They
have connected this changeability with depletion of the electron
neutrino flux passing through the SF region (hypothesis of the
$\nu_e$-induced $\beta$-decays). In Ref. \cite{BOG16} one was
supposed that this depletion may be bound by the neutrino
oscillations in the solar matter and solar magnetic field.
However, the analysis of that work has been fulfilled within two
flavor approximation. It might be well to point out that changing
the decay rate has been observed only for $\beta^{\pm}$ decay and
electron capture processes.

Neutrino oscillations in magnetic fields also allows to explain
the deficit of high-energetic muon neutrinos arising at long
Gamma-ray bursts (GRBs) which are probably connected with the
gravitational collapse of very massive stars. A black hole
produced during the collapse ejects two relativistic jets whose
magnetic fields could reach $10^8$ Gs. Besides producing
electromagnetic emission the GRBs could also be sources of cosmic
rays, neutrinos, and gravitational waves. In so doing the neutrino
energy could be as large as $10^{18}$ eV. There are a lot of works
devoted to studying the neutrino production in different scenarios
of GRBs. However, the upper limit on the high energy muon neutrino
flux obtained from the data collected with the 59-string
configuration of IceCube is 3.7 times below existing theoretical
predictions. It is not inconceivable that this decreasing may be
caused by neutrino resonance transitions as well \cite{FLV13}.

In the present work we shall continue investigation about behavior
of the solar neutrino flux which travels the region of the SF in
the preflare period. The investigation is carried out within the
context of three neutrino generations. The purpose of our work is
to answer the question whether it is possible to predict the SF's
observing the solar neutrino flux. In the next section we give a
brief sketch of the magnetic reconnection model which describes
the SF mechanism. The neutrino electromagnetic properties are
discussed in section 3. In section 4 we find the evolution
equation and define the possible resonance conversions of the
neutrino flux in the Sun's matter and magnetic field. Our
treatment of the problem carries rather general character, namely,
it holds for any standard model extensions in which neutrinos have
masses and possess both the magnetic dipole and anapole moments.
Section 5 is devoted to our conclusions. The natural system of
units ($\hbar=c=1$) is used.

\section{Magnetic reconnection model}
It is believed that the magnetic field is the main energy source
of the SF's. Note that one could only observe the magnetic
activity at the surface of the Sun and infer the magnetic field
inside. Therefore configuration and strength of the solar interior
magnetic field are not quite clear. However one may claim that in
the central part of the Sun's core, the magnetic field must not
exceed the value $B_c=5\times10^{7}$ Gs. Otherwise, as calculation
show, at $B>B_c$ this magnetic field would be lost by the Sun due
to the effect of "floating to the surface" during its existence.
Both in the center core and in the radiative zone the fields do
not display the time dependence. In the convective zone the
magnetic field module has a 11.2-yr cycle and in its bottom the
field could reach the value of $10^5$ Gs while its value at the
surface totaly depends on the existence of the AR's. During the
years of the active Sun, the magnetic flux $\sim10^{24}\
\mbox{Gs}\cdot\mbox{cm}^2$ \cite{DJ81} erupts from the solar
interior and accumulates to form the AR's. The flux collects
within the AR's giving rise to the stored magnetic field $B_s$. In
those places of the AR's where the magnetic field value reaches
500 Gs the process of producing sunspots begins. A typical size of
sunspots has the order of the Earth's radius
($R_{\oplus}=6.37\times10^8$ cm) in diameter and its hight may
reach the corona level. One could estimate the magnetic field
strength of sunspots which will be, for example, the source of the
super-SF's. If we assume that the magnetic field of such a sunspot
extends to the distance $h\simeq10^7$ cm and that the magnetic
energy stored with the volume $V=\pi R_{\oplus}^2h$ is equal to
$10^{36}$ erg, then we get $B_s\simeq\mbox{few}\times10^6$ Gs. In
fact, the value of $B_s$ must be greater, since only a small
portion of the total energy of a sunspot can be used, that is, a
large amount of energy is unavailable because it is distributed as
the potential field energy.

The magnetic field in the convective zone is characterized by the
geometrical phase $\Phi(z)$ defined by the relation
$$B_x\pm iB_y = B_{\bot}e^{\pm i\Phi(z)}\eqno(1)$$
and its first derivative on $z$, $\dot{\Phi}(z)$, in another way,
the magnetic field exhibits the twisting nature and has the twist
frequency. Nonzero values of $\Phi(z)$ and $\dot{\Phi}(z)$ also
exist in the photosphere and the chromosphere in regions above
sunspots. The magnetic field above and under sunspots has
the nonpotential character
$$(\mbox{rot}\ {\bf{B}})_z=4\pi j_z,\eqno(2)$$
where $j_z$ is the electric current density.
The data concerning centimeter radiation above a spot is
indicative of a gas heating up to the temperatures of a coronal
order. For example, at the height $\sim 2\cdot 10^2$ km the
temperature could be as large as $10^6$ K, that leads to a great
value of solar plasma conductivity ($\sigma\sim T^{3/2}$). That
permits to assume, that the longitudinal electric current $J_z$
might be large enough in a region above sunspots. In Ref.
\cite{KDA15} it was shown that when the magnetic field of newly
emerged sunspot takes the value 2000 Gs, $J_z$ can reach
$(0.7-4)\times10^{12}$ A. Then, for the sunspot with $R_s=10^8\
\mbox{cm}$ the electric current density ranges between $(0.7 -
4)\times10^{-1}\ \mbox{mA}/\mbox{cm}^2$.

The commonly accepted model of the SF production is the magnetic
reconnection model (MRM) which is based on breaking and
reconnection of magnetic field strength lines of neighboring
sunspots. This mechanism suggested in Ref. \cite{PA64} further on
was developed in details in Refs. {\cite{SI81,KST11}}. According
to the MRM the process of the SF evolution is as follows. The SF
formation starts from the integration of group of big sunspots in
pairs of opposite polarity (in what follows we shall call them
{\it{coupled sunspots}}). Then changing the magnetic field
configuration could result in the appearance of a limiting
strength line being common for the coupled sunspots. Throughout
this line which rises from photosphere to the corona the
redistribution of magnetic fluxes incoming from the solar interior
got under way. From the moment of appearance of the limiting
strength line, an electric field induced by magnetic field
variations causes current along this line. By virtue of the
interaction with a magnetic field this current takes the form of a
current layer (CL). Because the CL prevents from the magnetic
fluxes redistribution, the process of magnetic energy storage of
the CL begins. In so doing the magnetic field of the coupled
sunspots acquires the magnetic energy excess of the CL. The
greater the magnetic field of the coupled sunspots was, the
powerfuler the SF will be. The duration of the formation period of
the CL (the SF initial phase) varies from several to dozens of
hours. At this phase the magnetic field value for coupled sunspots
$B$ could be increased from $\sim10^4$ Gs up to $\sim10^5$ Gs and
upwards. The second SF stage (the explosion phase) at which the CL
is broken has a time interval of 1-3 minutes. The cause for the
rupture of the CL is thermal instability, which leads to the chain
of kinetic phenomena: (i) the rapid heating of plasma electrons;
(ii) the excitation of a plasma instability; and (iii) the
transition of the CL to a turbulent state. In that case the
electric resistance of the CL increases sharply. The appearance in
a certain part of the CL of a region of high or anomalous
resistance leads to the rapid current dissipation and,
accordingly, to the penetration of the magnetic fields through the
CL. The latter phenomenon is accompanied by a reconnection of the
magnetic field lines, which is why it has been called the magnetic
reconnection. A strong magnetic field arises across the CL, which
creates a magnetic force that tends to break the CL. Under the
action of this force, the plasma is ejected from the region of the
CL at high speed. The magnetic energy of coupled sunspots is
transformed into kinetic energy of matter emission (at a speed of
the order of $10^6\ \mbox{m/s}$), energy of hard electromagnetic
radiation, and fluxes of solar cosmic rays which consist of
protons, nuclei with charges $2\leq {\mbox{Z}}\leq28$, and
electrons. The produced photons reach the Earth by approximately
8.5 minutes after the explosion phase of the SF. Further during
some tens of minutes powerful flux of charged particles attains
terrestrial surface. As far as the plasma clouds are concerned,
they reach our planet within two-three days only. The most
powerful flux falling onto the Earth's surface may reach $\sim
4500\%$ in comparison to the background flux of cosmic particles.
The concluding SF stage (the hot phase) could continue for several
hours. It is exemplified by the existence of a high temperature
coronal region which consists of dense hot plasma cloud. One of
the characteristic features of flares is their isomorphism, that
is, the repetition in one and the same place with the same field
configuration. A small flare may repeat up to 10 times per day
while a large one may take place the next day and even several
times during the active region lifetime.

Note, there are some kinds of models which predict different
values for the magnetic reconnection rates at the explosion phase
of the SF. For discussing of this problem see, for example, Ref.
\cite{EPL}. However, in this
work our interest is in the investigation of the SF initial
phase only.

\section{Neutrino multipole moments}

In this section we shall discuss the neutrino electromagnetic
properties. Neutrinos are neutral particles and their total
Lagrangian does not contain any electromagnetic multipole moments
(MM's). These moments are caused by the radiative corrections
(RC's). The results of the RC's are usually reported in terms of
the effective Lagrangian
$${\cal{L}}_{em}={i\over2}\mu_{ll^{\prime}}\overline{\nu}_l(x)
\sigma^{\mu\lambda}(1-\gamma_5)\nu_{l^{\prime}}(x)F_{\lambda\mu}(x)+
{i\over2}a_{ll^{\prime}}\overline{\nu}_l(x)(\partial^{\mu}\gamma^{\lambda}-
\partial^{\lambda}\gamma^{\mu})(1-\gamma_5)
\nu_{l^{\prime}}(x)F_{\lambda\mu}(x)=$$
$$={i\over2}\mu_{ab}\overline{\nu}_a(x)
\sigma^{\mu\lambda}(1-\gamma_5)\nu_{b}(x)F_{\lambda\mu}(x)+
{i\over2}a_{ab}\overline{\nu}_a(x)(\partial^{\mu}\gamma^{\lambda}-
\partial^{\lambda}\gamma^{\mu})(1-\gamma_5)
\nu_{b}(x)F_{\lambda\mu}(x)+\mbox{conj.},\eqno(3)$$
where
the indexes $l,l^{\prime}$ refer to the flavor basis
($l,l^{\prime}=e,\mu,\tau$) while the indexes $a$ and $b$ refer to
the mass eigenstate basis ($a,b=1,2,3$), $\mu_{ab}$ ($a_{ab}$) are
the dipole magnetic (anapole) moments of the mass eigenstates, and
$F_{\lambda\mu}=\partial_{\lambda}A_{\mu}-\partial_{\mu}A_{\lambda}$.

For a Majorana neutrino from the $CPT$ invariance it is evident
that all the diagonal MM's, except the anapole one, are
identically equal to zero. As regards non-diagonal elements, the
situation depends on the fact whether $CP$-parity is conserved or
not. For the $CP$ non-variant case all MM's are nonzero. When $CP$
invariance takes place and the $\nu_{initial}$ and $\nu_{final}$ states
have identical (opposite) $CP$-parities, then $a_{ab}$
($\mu_{ab}$) are different from zero.

Further we address the experimental bounds on the dipole magnetic
and anapole neutrino moments. Let us start with the Dirac
neutrinos. The Borexino experiments give the limits on the DMM's
of the form \cite{DMP8,MAG17}
$$\mu_{\nu_e\nu_e}\leq2.9\times10^{-11}\mu_B,\qquad
\mu_{\nu_{\mu}\nu_{\mu}}\leq1.5\times10^{-10}\mu_B,\qquad
\mu_{\nu_{\tau}\nu_{\tau}}\leq1.9\times10^{-10}\mu_B,\eqno(4)$$
where $\mu_B$ is the Bohr magneton. As far as the bounds on
transit DMMs are concerned, they will be obtained only under
observation of processes proceeding with the partial lepton flavor
violation. In the case of Majorana neutrinos the global fit of the
reactor and solar neutrino data result in the following bounds for
transition DMMs \cite{WG03}
$$\mu_{12}, \mu_{13}, \mu_{23}\leq1.8\times10^{-10}\mu_B.\eqno(5)$$

The value of the anapole moment is connected with the charge
radius through the relation (see, for example, \cite{AR00})
$$a_{\nu_i}={1\over6}<r^2(\nu_i)>.\eqno(6)$$
The relation (6) is obtained within the SM and it is model
dependent. Moreover, even in the SM,
this relation is valid only
for massless neutrinos. It should be also recorded that, by now,
calculation of the anapole moment has been fulfilled only within
the SM in the case of both massless and massive Dirac neutrinos.
Therefore, it is not improbable that in the SM extension the
anapole moment value appears to be much bigger than that predicted
by the SM, as happened with the DMM's. Remember, the DMM values in
the SM are given by the expression \cite{BWLe77}
$$\mu_{\nu_l\nu_l}=10^{-19}\ \mu_B \Bigg({m_{\nu_l}
\over\mbox{eV}}\Bigg),\eqno(7)$$ while in models containing
right-handed charged currents and/or charged Higgs bosons
$\mu_{\nu_l\nu_l}$ is proportional to the charged lepton
mass $m_l$ and
proves to be on 7-8 orders of magnitude bigger (see, for example,
Ref. \cite{OM9014}).

One should remember that the right dimensionality of the anapole
moment in CGS system is "$\mbox{length}^2\times\mbox{charge}$"
\cite{IaBZ}). So, to turn from the natural system of units to CGS
system the $a_{\nu_i}$ value must be multiplied by $\sqrt{\hbar
c}$.

Measuring the elastic neutrino-electron scattering at the TEXONO
experiment leads to the following bounds on the electron neutrino
charge radius (ENCR) \cite{TSK15}
$$-2.1\times10^{-32}\ \mbox{cm}^2\leq<r_{\nu_e}^2>\leq
3.3\times10^{-32}\ \mbox{cm}^2.\eqno(8)$$ There are other limits
on the ENCR as well. They are derived from neutrino
neutral-current  reactions \cite{ANK17}
$$-2.74\times10^{-32}\mbox{cm}^2\leq<r_{\nu_e}^2>\leq
4.88\times10^{-32}\ \mbox{cm}^2.\eqno(9)$$ Calculations carried
out within the SM \cite{DSM89} lead to the conclusion that the
charge radiuses of $\nu_{eL},\nu_{\mu L}$ and $\nu_{\tau L}$ have
the same order, namely, $\mbox{few}\times10^{-32}\ \mbox{cm}^2$.
However, it must be emphasized that the boundaries (8) and (9)
were obtained under comparison of experimental results with the
theoretical expressions for the corresponding cross sections
obtained within the SM. Since similar analysis was not completed
with alternative models, we have to use the above mentioned
boundaries for the ENCR.

Further, making numerical estimates, we shall take the following
values for the MM's neutrino
$$\mu_{\nu_l\nu_{l^{\prime}}}=10^{-10}\ \mu_B,
\qquad |a_{\nu_l\nu_{l^{\prime}}}|=3\times10^{-40}\
\mbox{esu}\cdot\mbox{cm}^2,$$ where esu (electrostatic unit) is
the unit of measurement of electricity in the CGS system. As for
the magnetic field of the coupled sunspots, we shall assume that
$B_s\geq10^5$ Gs.

\section{Solar neutrino flux}
We are coming now to the analysis of the evolution equation of the
neutrino flux traveling the SF region. We shall work within the three
neutrino generations. In so doing we are going to allow for
interaction not only with solar matter, but with solar magnetic
field as well. Therefore, the system under study must include both
the left-handed and right-handed neutrinos, that is, its wave function
must be as follows $\psi^T=(\nu_{eL},\nu_{\mu L},\nu_{\tau L},\nu_{eR},
\nu_{\mu R},\nu_{\tau R})$.
For the magnetic field of coupled sunspots we shall adopt a simple
model in which
$$\Phi(z)={\alpha\pi\over L_{mf}}z,\eqno(10)$$
that is, the magnetic field exists over a distance $L_{mf}$ and
twists by an angle $\alpha\pi$ ($\alpha\pi/L_{mf}$ is the twist
frequency).

The current values of oscillation parameters we are interested in
are as follows \cite{FC17}
$$\left.\begin{array}{lll}\Delta m^2_{31(23)}\simeq2.56\times10^{-3}\
\mbox{eV}^2,\hspace{4mm} \Delta m^2_{21}\simeq7.87\times10^{-5}\
\mbox{eV}^2,\hspace{4mm} \sin^2\theta_{12}\simeq0.297,\\ [2mm]
\sin^2\theta_{13}\ (\Delta m_{31(32)}>0)\simeq0.0215, \qquad
\sin^2\theta_{13}\ (\Delta m_{31(32)}<0)\simeq0.0216,\\ [2mm]
\sin^2\theta_{23}\ (\Delta m_{31(32)}>0)\simeq0.425, \qquad
\sin^2\theta_{23}\ (\Delta m_{31(32)}<0)\simeq0.589.
\end{array}\right\}\eqno(11)$$

In order to get the evolution equation we shall use the standard
technique of obtaining the similar equations (see, for example,
the books \cite{OMB11,KZ11}). The basic idea of this approach
consists in the reduction of the totality of the neutrino
interactions in matter and magnetic field to the motion in a field
with a potential energy. As this takes place, to find the matter
potential one should first consider the neutrino interactions with
single electron, neutron, proton and then fulfill averaging over
all matter particles. Taking into account Eg.(3) and assuming the
Dirac neutrino nature we obtain the required equation
$$i{d\over dz}\left(\matrix{\nu_{eL}\cr\nu_{\mu L}\cr\nu_{\tau
L}\cr\nu_{eR}\cr\nu_{\mu R}\cr\nu_{\tau R}}\right)={\cal{H}}
\left(\matrix{\nu_{eL}\cr\nu_{\mu L}\cr\nu_{\tau
L}\cr\nu_{eR}\cr\nu_{\mu R}\cr\nu_{\tau R}}\right),\eqno(12)$$ where
$${\cal{H}}={\cal{U}}\left(\matrix{E_1&0&0&\mu_{11}B_{\perp}e^{i\Phi}&
\mu_{12} B_{\perp}e^{i\Phi}&\mu_{13}B_{\perp}e^{i\Phi}\cr
0&E_2&0&\mu_{21}B_{\perp}e^{i\Phi}&\mu_{22}B_{\perp}e^{i\Phi}&\mu_{23}
B_{\perp}e^{i\Phi}
\cr0&0&E_3&\mu_{31}B_{\perp}e^{i\Phi}&\mu_{32}B_{\perp}e^{i\Phi}&\mu_{33}
B_{\perp}e^{i\Phi}e^{i\Phi}
\cr\mu_{11}B_{\perp}e^{-i\Phi}&\mu_{12}B_{\perp}e^{-i\Phi}&\mu_{13}
B_{\perp}e^{-i\Phi}&E_1&0&0\cr
\mu_{21}B_{\perp}e^{-i\Phi}&\mu_{22}B_{\perp}e^{-i\Phi}&\mu_{23}
B_{\perp}e^{-i\Phi}&0&E_2&0 \cr\mu_{31}B_{\perp}e^{-i\Phi}
&\mu_{32}B_{\perp}e^{-i\Phi}&\mu_{33}B_{\perp}e^{-i\Phi}&0&0&E_3}\right)
{\cal{U}}^{-1}+$$
$$+\left(\matrix{V_{eL}+d_{\nu_{eL}\nu_{eL}}&0&0&0&0&0 \cr0&V_{\mu L}+
d_{\nu_{\mu L}\nu_{\mu L}}&0&0&0&0\cr 0&0&V_{\tau L}+d_{\nu_{\tau L}\nu_{\tau L}}
&0&0&0\cr 0&0&0&-d_{\nu_{eR}\nu_{eR}}&0&0\cr 0&0&0&0&-d_{\nu_{\mu R}\nu_{\mu R}}
&0\cr0&0&0&0&0&-d_{\nu_{\tau R}\nu_{\tau R}}}
\right),$$
$${\cal{U}}=\left(\matrix{\cal{D}& 0\cr 0&\cal{D}}\right),$$
$${\cal{D}}=\exp({i\lambda_7\psi})\exp({i\lambda_5\phi})
\exp({i\lambda_2\omega})=\left(\matrix{c_{\omega}c_{\phi} &
s_{\omega}c_{\phi} & s_{\phi}\cr
-s_{\omega}c_{\psi}-c_{\omega}s_{\psi}s_{\phi} &
c_{\omega}c_{\psi}-s_{\omega}s_{\psi}s_{\phi} &
s_{\psi}c_{\phi}\cr s_{\omega}s_{\psi}-c_{\omega}c_{\psi}s_{\phi}
& -c_{\omega}s_{\psi}-s_{\omega}c_{\psi}s_{\phi}&
c_{\psi}c_{\phi}}\right),$$ $\psi=\theta_{23}, \ \phi=\theta_{13},
\ \omega=\theta_{12},$ $s_{\psi}=\sin\psi, c_{\psi}=\cos\psi,
s_{\phi}=\sin\phi, c_{\phi}=\cos\phi, s_{\omega}=\sin\omega,
c_{\omega}=\cos\omega$,
the $\lambda$'s are Gell-Mann matrices
corresponding to the spin-one matrices of the $SO(3)$ group,
$V_{eL}$ ($V_{\mu L}$) is a matter potential describing
interaction of the $\nu_{eL}$ ($\nu_{\mu L},\nu_{\tau L}$)
neutrinos with a solar matter,
$$V_{eL}=\sqrt{2}G_F(n_e-n_n/2),\qquad
V_{\mu L}=V_{\tau L}=-\sqrt{2}G_Fn_n/2,$$ $n_e$ and $n_n$ are
electron and neutron densities, respectively,
$$d_{\nu_{lL}\nu_{lL}}=4\pi a_{\nu_{lL}\nu_{lL}}j_z,\qquad
d_{\nu_{lR}\nu_{lR}}=4\pi a_{\nu_{lR}\nu_{lR}}j_z,$$
$a_{\nu_l\nu_{l^{\prime}}}$ ($\mu_{\nu_l\nu_{l^{\prime}}}$) is an
anapole (dipole magnetic) moment between $\nu_l$ and
$\nu_{l^{\prime}}$ states, and, for the sake of simplicity, we
have assumed that the nondiagonal neutrino anapole moments are
equal to zero.

In Eq.(12) one should get rid of imaginary part in Hamiltonian. It
is achieved by transformation to reference frame (RF), rotating at
the same angle speed as a magnetic field \cite{AYU91}.
The matrix of
transition to the new RF will look like
$$S=\mbox{diag}(\lambda,\lambda,
\lambda,-\lambda,-\lambda,-\lambda),\eqno(13)$$ where
$\lambda=\exp{(i\Phi/2)}.$ The Hamiltonian in this RF follows from
the initial one by a replacement
$$e^{\pm i\Phi}\longrightarrow 1,\qquad
4\pi a_{\nu_{lL}\nu_{lL}}j_z\to4\pi a_{\nu_{lL}\nu_{lL}}j_z-\dot{\Phi}/2,
\qquad 4\pi a_{\nu_{lR}\nu_{lR}}j_z\to 4\pi
a_{\nu_{lR}\nu_{lR}}j_z-\dot{\Phi}/2.\eqno(14)$$

In general, the evolution equation (12) could be solved
numerically or with appropriate approximations. In our case to
define all possible electron neutrino resonance conversions in the
system under study and make the results physically more
transparent, one may proceed in the following manner. We shall
search for such a basis in which, on the one hand, physical
implications will be evident, and, on the other hand, one of the
states will be predominantly the $\nu_{eL}$ state. Taking into
account smallness of the mixing angle $\phi$ we find the required
transformation
$$\left(\matrix{\nu_{1L}\cr\nu_{2L}\cr\nu_{3L}
\cr\nu_{1R}\cr\nu_{2R}\cr
\nu_{3R}}\right)={\cal{U}}^{\prime}
\left(\matrix{\nu_{eL}\cr\nu_{\mu L}\cr\nu_{\tau
L}\cr\nu_{eR}\cr\nu_{\mu R}\cr\nu_{\tau R}}\right),\eqno(15)$$ where
$${\cal{U}}^{\prime}=\left(\matrix{{\cal{D}}^{\prime}& 0\cr
0&{\cal{D}}^{\prime}}\right),\qquad {\cal{D}}^{\prime}=
\exp({-i\lambda_5\phi})\exp({-i\lambda_7\psi})=
\left(\matrix{c_{\phi} &0 &s_{\phi}\cr -s_{\phi}s_{\psi} &
c_{\psi}& c_{\phi}s_{\psi}\cr-s_{\phi}c_{\psi}& -s_{\phi}&
c_{\phi}c_{\psi}}\right).\eqno(16)$$
From (15) it follows that the $\nu_{1L}$ ($\nu_{3L}$)
state is predominately the $\nu_{eL}$ ($\nu_{\tau L}$) flavor
state while the $\nu_{2L}$ state represents the mixing of
the $\nu_{\mu L}$ and $\nu_{\tau L}$ flavor states. The same is
true for their corresponding right-handed partners.

The transformed Hamiltonian acquires the form
$${\cal{H}}^{\prime}={\cal{U}}^{\prime}\ {\cal{H}}\
{\cal{U}}^{\prime-1}=\left(\matrix{{\cal{B}}_v+ \Lambda
&\cal{M}\cr\cal{M}&{\cal{B}}_v+\tilde{\Lambda}}\right),\eqno(17)$$
where
$${\cal{B}}_v=\left(\matrix{-\delta^{12}c_{2\omega} & \delta^{12}
s_{2\omega}&0\cr
\delta^{12}s_{2\omega}&\delta^{12}c_{2\omega}&0\cr
0&0&\delta^{31}+\delta^{32}}\right),\qquad
\Lambda=\left(\matrix{V_{eL}^{eff}c_{\phi}^2& 0&
V_{eL}^{eff}s_{2\phi}/2\cr 0&0&0\cr
V_{eL}^{eff}s_{2\phi}/2&0&V_{eL}^{eff}s_{\phi}^2}\right),$$
$$\tilde{\Lambda}=\left(\matrix{{\cal{A}}_{\nu_{lR}\nu_{lR}}-
{\cal{A}}_{\nu_{lL}\nu_{lL}}-V_{\mu L}&0&0\cr 0&
{\cal{A}}_{\nu_{lR}\nu_{lR}}-
{\cal{A}}_{\nu_{lL}\nu_{lL}}-V_{\mu
L}&0\cr0&0&{\cal{A}}_{\nu_{lR}\nu_{lR}}-
{\cal{A}}_{\nu_{lL}\nu_{lL}}-V_{\mu L}}\right),$$
$${\cal{M}}=\left(\matrix{(\mu_0+\mu_{12}s_{2\omega})B_{\perp}
&\mu_{12}c_{2\omega}B_{\perp}&(\mu_{13}c_{\omega}+\mu_{23}s_{\omega})
B_{\perp}\cr \mu_{12}c_{2\omega}B_{\perp}&
\mu_{12}c_{2\omega}B_{\perp}&
(\mu_{13}c_{\omega}+\mu_{23}s_{\omega})B_{\perp}\cr
0&0&\mu_{33}B_{\perp}}\right),$$
$$\delta^{ik}={m_i^2-m_{k}^2\over4E},\qquad \mu_{11}=\mu_{22}=\mu_0/2,
\qquad V_{eL}^{eff}=\sqrt{2}G_Fn_e,$$
$${\cal{A}}_{\nu_{lL}\nu_{lL}}=4\pi a_{\nu_{lL}\nu_{lL}}j_z-\dot{\Phi}/2,\qquad
{\cal{A}}_{\nu_{lR}\nu_{lR}}=-4\pi a_{\nu_{lR}
\nu_{lR}}j_z+\dot{\Phi}/2,$$
and, for the sake of simplicity, we have set
$$a_{\nu_{eL}\nu_{eL}}=a_{\nu_{\mu L}\nu_{\mu L}}=a_{\nu_{\tau L}\nu_{\tau L}}=
a_{\nu_{lL}\nu_{lL}},\qquad
a_{\nu_{eR}\nu_{eR}}=a_{\nu_{\mu R}\nu_{\mu R}}
=a_{\nu_{\tau R}\nu_{\tau R}}=a_{\nu_{lR}\nu_{lR}}.$$

Now, using the expression for ${\cal{H}}^{\prime}$, we can
establish all possible resonance conversions. We shall assume that the
resonance localization places
are situated rather far from one another. That allows us to
consider them as independent ones. We shall also
be constrained by consideration of the resonance transitions with the
participation of the $\nu_{1L}$ neutrino only.

Let us start with the
$\nu_{1L}\to\nu_{2L}$ transition. Equating the
corresponding diagonal elements of the Hamiltonian
${\cal{H}}^{\prime}$ we obtain the conditions of the resonance
existence (in what follows we shall use the term "resonance condition")
$$-2\delta^{12}c_{2\omega}+V^{eff}_{eL}c_{\phi}^2=0.\eqno(18)$$ To deeper realize
consequences of neutrino behavior we proceed as follows.
We infer that the matter density is
constant. Then the expression for the transition probability of
the neutrino system consisting only from $\nu_{1L}$ and
$\nu_{2L}$ will look like
$$P_{\nu_{1L}\to\nu_{2L}}(z)\simeq\sin^22\theta_m
\sin^2\Bigg({z\over
L_{\nu_{1L}\nu_{2L}}}\Bigg),\eqno(19)$$
where
$L_{\nu_{1L}\nu_{2L}}$ is the oscillation length of
the $\nu_{1L}\to\nu_{2L}$ resonance
$$L_{\nu_{1L}\nu_{2L}}={2\pi\over\sqrt{
[2\delta^{12}{\displaystyle{c_{2\omega}}}
-V_{eL}^{eff}\displaystyle{c_{{\phi}}^2}]^2+(2
\delta^{12}{\displaystyle{s_{2\omega}}})^2}},\eqno(20)$$
$\theta_m$ is a mixing angle in a matter
$$\tan2\theta_m={2{\cal{H}}^{\prime}_{\nu_{1L}\nu_{2L}}\over
{\cal{H}}^{\prime}_{\nu_{2L}\nu_{2L}}-
{\cal{H}}^{\prime}_{\nu_{1L}\nu_{1L}}}
\simeq{2\delta^{12}s_{2\omega}\over2\delta^{12}c_{2\omega}-
V_{eL}^{eff}c_{\phi}^2}.\eqno(21)$$
The behavior character of the
mixing angle $\theta_m$ becomes more evident when we rewrite the
relation (21) in the form
$$\sin^22\theta_m={(2\delta^{12}s_{2\omega})^2\over
[2\delta^{12}c_{2\omega}-V^{eff}c_{\phi}^2]^2
+(2\delta^{12}s_{2\omega})^2}.\eqno(22)$$
From Eq.(22) it
immediately follows, in a solar matter with a variable electron
density the dependence of the mixing angle $\theta_m$ on $n_e$ has
a resonance character. When the condition (18) is fulfilled
$\theta_m$ reaches its maximum value $\pi/4.$ However, from
Eq.(19) it follows that for oscillations to be appeared a neutrino
beam must pass a distance comparable with oscillation length. Note
the oscillation length reaches its maximal value at the
resonance.

One more important characteristic of the resonance represents the
transition width. If it is equal to zero the resonance transition
will be forbidden, even though the resonance condition
is satisfied. For the $\nu_{1L}\to\nu_{2L}$
resonance it is given by the expression
$$\Gamma(\nu_{1L}\to\nu_{2L})\simeq
{\sqrt{2}\delta^{12}s_{2\omega}\over G_F}.\eqno(23)$$
Note that the expressions (19) - (23) coincide with the corresponding ones
describing the $\nu_{eL}\to\nu_{\mu L}$ resonance in two flavor approximation
(so called Micheev-Smirnov-Wolfenstein --- MSW resonance)
under substitution
$$n_e\to n_ec_{\Phi}^2.\eqno(24)$$

When we set
$$E=10\ \mbox{MeV},\qquad \Delta m^2=7.37\times10^{-5}\
\mbox{eV}^2,\qquad\ \sin^2\theta_{12}=0.297,\eqno(25)$$ then the maximum
oscillation length takes the value
$\simeq3.5\times10^7$ cm and, as a result, this resonance occurs before
the convective zone. Therefore it happens whether the SF being at work or
not. Since the expressions (18) - (22)
depend on the neutrino energy then only electron neutrinos
with the energy of order of few MeV take part in this resonance transition.

If one assumes that not only the matter density is a constant, but the
quantities $B_{\perp}$ and $j_z$ are constants as well, then, for the
well-separated resonances the transition probabilities will be given by
the expressions being analogous to (19) with corresponding values of the oscillation
length and the mixing angle.
It is obvious that in the real case, when we deal with the variable matter
density and variable magnetic field, the occurrence of a resonance will
be also dependent on values of such characteristics as
the resonance condition, the resonance width and the distance traveled by the
neutrino beam. By these reasons when discussing the resonances we shall be
limited by the analysis of these characteristics only.

We now turn to the discussion of the helicity flip resonance transitions.
For the first time within one-flavor approximation the existence of
$\nu_{lL}\to\nu_{lR}$ resonance
in a magnetic field was indicated in the work \cite{VVO}. In the literature,
this helicity flip transition (HFT) is often referred to as the
Voloshin-Vysotskii-Okun effect. Later the HFT's of the neutrino systems
traveling in magnetic fields were generalized to the case of two flavor
approximations (see, for example, \cite{EKh,XSH,OBD95,PAL}).
We start our consideration with the
$\nu_{1L}\to\nu_{2R}$ resonance.
It may be realized at the condition
$$-2\delta^{12}c_{2\omega}+V_{eL}c_{\phi}^2+4\pi(a_{\nu_{lL}\nu_{lL}}+
a_{\nu_{lR}\nu_{lR}})j_z-\dot{\Phi}=0.\eqno(26)$$
The corresponding expressions for the transition width and the
maximum oscillation length are as follows
$$\Gamma(\nu_{1L}\to\nu_{2R})\simeq
{\sqrt{2}\mu_{12}c_{2\omega}B_{\perp}\over G_F},\eqno(27)$$
$$(L_{\nu_{1L}\to\nu_{2R}})_{max}\simeq{2\pi\over
\mu_{12}c_{2\omega}B_{\perp}}.\eqno(28)$$

It should be pointed out that the sunspots occur
not only at the surface of the Sun, but at the convective zone as
well. Thanks to the Wilson depression \cite{PM01} they could lie
below the photosphere on $L_W\sim500 - 700$ km where the matter
potential $V_{eL}$ is nothing more than $\mbox{few}\times10^{-17}$
eV. One may assume that the preflare pairing of the Wilson's
sunspots takes place as well. Then the
$\nu_{1L}\to\nu_{2R}$ resonance may occur when the
neutrino beam passes through the magnetic field of these
coupled sunspots. However, we shall not consider the case of
Wilson's sunspots and assume that we deal with the coupled
sunspots positioned on the solar atmosphere.

For the solar neutrinos $(\delta^{12})_{min}\simeq10^{-12}$ eV
which is much more bigger than the matter potential even in
photosphere ($V_{ph}\simeq10^{-20}$\ eV).
Therefore, in the Sun's conditions the resonance
$\nu_{1L}\to\nu_{2R}$ may occur only at the
cost of magnetic field, that is, when the sum
$$2\delta^{12}c_{2\omega}+\dot{\Phi}-4\pi(a_{\nu_{lL}\nu_{lL}}+
a_{\nu_{lR}\nu_{lR}})j_z\eqno(29)$$ has the same order as
$V_{eL}c^2_{\phi}$.
So, we may say that this resonance falls to the kind
of the magnetic-induced resonances.
It should be emphasized that the magnetic field must be twisted and/or has
a nonpotential character.

Since the resonance condition (26) does not contain the value of $B_{\perp}$,
it may seem that the $\nu_{1L}\to\nu_{2R}$ resonance can
also occur when $B_{\perp}=0$. However,
it is not the case. Indeed, the resonance condition is valid also for tiny values of $B_\perp$ and
$\theta_m\to \pi/4$ for any value of $B_\perp$.
However, the oscillation length tends to infinity in the limit of $B_\perp\to
0$ making the transition impractical.

Comparing the expressions (26) - (28) with the corresponding ones
describing the $\nu_{eL}\to\nu_{\mu R}$ resonance in two flavor approximation
(FA) \cite{BOG16} one could be convinced that the formulas for three neutrino generations
are evident from those of two FA under substitutions
$$n_e\to n_ec_{\Phi}^2,\eqno(30)$$
$$\mu_{\nu_{eL}\nu_{\mu R}}\to
\mu_{12}c_{2\omega}.\eqno(31)$$

Using $B=10^5$ Gs we get
$(L_{\nu_{1L}\to\nu_{2R}})_{max}\simeq7\times10^8$
cm. Then the resonance condition and the equality
$L_{mf}=(L_{\nu_{1L}\to\nu_{2R}})_{max}$ will be
fulfilled provided the twist frequency $\dot{\Phi}$ is equal to
$-10\pi/L$, where we have assumed that
$$\dot{\Phi}\gg4\pi(a_{\nu_{lL}\nu_{lL}}+
a_{\nu_{lR}\nu_{lR}})j_z.$$ On the other hand
when the magnetic field reaches the
value of $10^6$ Gs what will be possible for the super-SF's, the
fulfillment above mentioned requirements will be effected at the
twist frequency being equal to $-\pi$ and $L_{mf}=7\times10^7$ cm.
So, we see that under the specific conditions the
$\nu_{1L}\to\nu_{2R}$ resonance may be in
existence.

The next resonance conversion is
$\nu_{1L}\to\nu_{1R}$.
The corresponding formulas will look like
$$V^{eff}_{eL}c_{\phi}^2+4\pi(a_{\nu_{lL}\nu_{lL}}
+a_{\nu_{lR}\nu_{lR}})j_z-\dot{\Phi}=0,\eqno(32)$$
$$\Gamma(\nu_{1L}\to\nu_{1R})\simeq
{\sqrt{2}(\mu_{11}+\mu_{12}s_{2\omega})B_{\perp}\over G_F},\eqno(33)$$
$$(L_{\nu_{1L}\to\nu_{1R}})_{max}
\simeq{2\pi\over(\mu_{11}+\mu_{12}s_{2\omega} )B_{\perp}}.\eqno(34)$$ It is
clear that the situation when $V^{eff}_{eL}c_{\phi}^2=\dot{\Phi}$ is
excluded, since in this case the twisting magnetic field must
exist over the distance which is much more even the solar radius.
So, only when the requirements
$$\dot{\Phi}\ll V^{eff}_{eL},\qquad\mbox{but}\qquad
V^{eff}_{eL}c_{\phi}^2\simeq-4\pi(a_{\nu_{lL}\nu_{lL}}+
a_{\nu_{lR}\nu_{lR}})j_z\eqno(35))$$ will be fulfilled the
$\nu_{1L}\to\nu_{1R}$ resonance may be observed.
Calculations demonstrate that all the expressions which govern the
$\nu_{1L}\to\nu_{1R}$ resonance conversion
may be deduced from the expressions for
$\nu_{eL}\to{\nu_{eR}}$ resonance obtained in  two FA
\cite{BOG16}
provided the replacement
$$n_e\to n_ec_{\Phi}^2,\eqno(36)$$
$$\mu_{\nu_{eL}\nu_{eR}}\to
\mu_{11}+\mu_{12}s_{2\omega}.\eqno(37)$$

We see that, in point of fact, the resonance condition
(32) is the distance function. On the other hand, since both the
resonance condition and the transition width do not display the
dependence on the neutrino energy, then all the electron neutrinos
produced in the center of the Sun ($pp$-, $^{13}N$-,...and
$hep$-neutrinos) may undergo $\nu_{1L}\to\nu_{1R}$
resonance transition.

Let us estimate the value of $j_z$ which is needed to realize the
$\nu_{1L}\to\nu_{1R}$ resonance in the
chromosphere (corona). Taking into account $n_n\simeq n_e/6$ we
obtain the following value for the matter potential $\sim10^{-27}$
eV ( $\sim10^{-30}$ eV). Further, assuming
$a_{\nu_{lL}\nu_{lL}}=a_{\nu_{lR}\nu_{lR}}$,
we see that the resonance condition (32) will be fulfilled
provided
$$j_z\simeq6\ \mbox{A}/\mbox{cm}^2\qquad (j_z\simeq0.06
\ \mbox{A}/\mbox{cm}^2).\eqno(38)$$

In what follows we are coming to consideration of the resonance
conversions which are absent in two FA, namely, to the
$\nu_{1L}\to\nu_{3L}$ and
$\nu_{1L}\to\nu_{3R}$ transitions. For the Sun
conditions the relation
$$V^{eff}_{eL},\Delta m^2_{12}\ll\Delta m^2_{23},\Delta
m^2_{13}\eqno(39)$$ holds. The quantity proportional to
$\Sigma=\delta^{31}+\delta^{32}$ is the dominant term in the
Hamiltonian (17) and this leads to the decoupling of $\nu_{3L}$
from the remaining states apart from the
$\nu_{3R}$ one. This means that the
oscillation $\nu_{1L}\to\nu_{3L}$ which is driven by
the $\Sigma$ term can be simply averaged out in the final survival
probability of electron neutrinos at the Earth.

As far as the $\nu_{1L}\to\nu_{3R}$
resonance is concerned, the situation here is not so
obvious and requires a more detailed analysis. The resonance condition,
the transition width and
maximum oscillation length are given by the expressions
$$V_{eL}^{eff}c_{\Phi}^2+V_{\mu L}+4\pi
(a_{\nu_{lL}\nu_{lL}}+a_{\nu_{lR}\nu_{lR}})j_z
-\delta^{12}c_{2\omega}-\Sigma-\dot{\Phi}=0,\eqno(40)$$
$$\Gamma(\nu_{1L}\to\nu_{3R})\simeq
{\sqrt{2}(\mu_{13}c_{\omega}+\mu_{23}s_{\omega})B_{\perp}\over G_F},\eqno(41)$$
$$(L_{\nu_{1L}\to\nu_{3R}})_{max}\simeq{2\pi\over
(\mu_{13}c_{\omega}+\mu_{23}s_{\omega})B_{\perp}}.\eqno(42)$$
At the first glance it would seem that the quantities
$\dot{\Phi}$ and $4\pi
(a_{\nu_{lL}\nu_{lL}}+a_{\nu_{lR}\nu_{lR}})j_z$
could cancel the big value of $\Sigma$. However, such is not
the case. So, for example, requiring the fulfillment
$\dot{\Phi}\simeq\Sigma,$ even when $B_{\perp}=10^6$ Gs, we get
$$L_{mf}\ll(L_{\nu_{1L}\to\nu_{3R}})_{max}.
$$ Therefore, in the Sun conditions the $\nu_{1L}\to\nu_{3R}$
resonance proves to be forbidden.

With a knowledge of the transition probabilities ${\cal{P}}(\nu_{1L}\to
\nu_{2L})$, ${\cal{P}}(\nu_{1L}\to
\nu_{1R})$, ${\cal{P}}(\nu_{1L}\to
\nu_{2R})$ and taking into consideration the flavor contents of the
$\nu_{lL}$ and $\nu_{lR}$ states, we could find
the electron neutrino survival probability
$${\cal{P}}(\nu_{eL}\to\nu_{eL})=1-c_{\phi}^2\Big[{\cal{P}}(\nu_{1L}
\to\nu_{2L})+{\cal{P}}(\nu_{1L}
\to\nu_{1R})+
{\cal{P}}(\nu_{1L}
\to\nu_{2R})\Big]+$$
$$+s_{\phi}^4s_{\psi}^2{\cal{P}}(\nu_{1L}
\to\nu_{2R}).\eqno(43)$$
Further we assume that the transition probabilities depend only on the mixing angles
and the oscillation lengths, as happens with the constant values of $n_e$, $j_z$
and $\dot{\Phi}$. When in (43) we
make any allowance for the connection between
$\mu_{ab}$ and $\mu_{ll^{\prime}}$ (see Eqs. (31) and (37), put
$\phi$ and $\psi$ equal to zero, then,
as would be expected, the expression (43) converts to the survival probability
for the electron neutrino in two FA.

Note that the majority of resonances have an energy range in which
neutrino conversion occurs. Since any given experiment is only
sensitive to a small, finite range of energies, it will generally
overlap only one of the transition regions.

It should be stressed that since the transition width of the MSW
resonance does not depend on the DMM then it proves to be allowed
within the SM. As far as the remaining magnetic-induced
resonances are concerned, their realization is possible only in the
model with nonzero DMM.

\section{Conclusions}

The goal of this work was to investigate the influence of the
solar flares (SF's) on behavior of solar neutrino fluxes. Within three
neutrino generations the evolution of the neutrino flux traveling
the coupled sunspots (CS's) being the SF source has been studied.
One was assumed that the neutrinos possess both the dipole
magnetic moment and the anapole moment while the magnetic field
above the CS's has the twisting nature and displays the nonpotential
character. We also inferred that in the process of magnetic energy
storage the strength of this field may reach the values of
$10^5-10^6$ Gs. For the analysis of the evolution equation we have transferred
to the new basis
in which one of the states $\nu_{1L}$ was predominantly the $\nu_{eL}$ state
($\nu_{eL}=\nu_{1L}\big|_{\phi=0}$).
This permits to connect the evolution of the electron neutrino beam with the behavior
of the $\nu_{1L}$ state. The possible resonance conversions with the
participation of the $\nu_{1L}$ neutrino have been examined.

Since the $\nu_{eL}\to\nu_{\mu}$ resonance ( MSW resonance) occurs before
the convective zone, its existence can in no way be connected with the SF.
After escaping
the Sun, the neutrino flux flies $1.5\times10^8$ km in a vacuum
before it will attain a terrestrial observer. In so doing
reduction of the electron neutrino flux is caused by the vacuum
oscillations which brings about $\nu_{eL}\to\nu_{\mu L}$
transitions only. Therefore, when the SF is absent, the neutrino
telescopes detect the electron neutrino flux weakened at the
expense both of vacuum oscillations and of the MSW resonance.
However, when the electron neutrino flux travels the magnetic
field of the CS's then it may be further weakened because of
additional resonance conversions, apart from the above-listed. In
the case of Dirac neutrinos the following resonances
$\nu_{eL}\to\nu_{eR}$, $\nu_{eL}\to\nu_{\mu R}$ and $\nu_{eL}\to\nu_{\tau R}$
could take place.

The conditions of the resonances existence and the transition
widths (TW's) have been found. It is worth noting that since for
the $\nu_{eL}\to\nu_{eR}$ resonance both the resonance
condition and the TW do not depend on the neutrino energy then all
electron neutrino born in the Sun's center may go through the
$\nu_{eL}\to\nu_{eR}$-resonance. The TW's of the
resonances $\nu_{eL}\to\nu_{eR}$,
$\nu_{eL}\to\nu_{\mu R}$, and
$\nu_{eL}\to\nu_{\tau R}$ proves to be proportional to
the neutrino dipole magnetic moment (DMM). Since the standard
model (SM) predicts the neutrino DMM value close to zero, then
from the SM point of view these resonances are forbidden.

So, under passage of the electron neutrino flux through the region
of the SF one may observe the depletion of the electron neutrino flux.
If the hypothesis of the
$\nu_{eL}$-induced $\beta$-decays is valid then observation of changeability
of the $\beta$-decay rates of some
elements during the SF's may be viewed as experimental
confirmation of decreasing the solar neutrino flux. Needles to say the
existence of such depletion must
be confirmed by other experiments. It could be done at the
neutrino telescopes of the next generation in which the events
statistics will be increased on several orders of magnitude (for
example, at the Fermi Lab Liquid ARgon experiment --- FLARE).

In summary, we emphasize that the conditions for emergence of the
$\nu_{eL}\to\nu_{eR}$, $\nu_{eL}\to\nu_{\mu R}$, $\nu_{eL}\to\nu_{\tau R}$ resonances contains two
uncertainties, namely, the value of the magnetic field above the
CS's providing the SF source, and the values of the neutrino
multipole moments. Therefore, knowledge of these parameters will
allow us to give the ultimate answer, whether it is possible or
not to predict the SF's by observing solar neutrino fluxes.

\section*{Acknowledgments}
This work is partially supported by the grant of Belorussian
Ministry of Education No 20162921.

\end{document}